\documentclass[preprintnumbers, pra, showpacs, floatfix,twocolumn,
preprintnumbers, letterpaper, superscriptaddress]{revtex4}
\usepackage{float}
\usepackage{amsfonts}
\usepackage{amsmath}
\usepackage{amssymb,epsf}
\usepackage{latexsym}
\usepackage{graphicx,epsfig}
\usepackage{amssymb}
\usepackage{subfigure}
\usepackage[colorlinks=true,citecolor=blue,linkcolor=blue,urlcolor=black]{hyperref}
\usepackage{epstopdf}
\usepackage{color}

\def\be{\begin{equation}}
\def\ee{\end{equation}}
\def\ba{\begin{eqnarray}}
\def\ea{\end{eqnarray}}

\def\ra{\rangle}

\begin{document}
\title{Layer-by-layer disentangling two-dimensional topological quantum codes}
\author{Mohammad Hossein Zarei}
\email{mzarei92@shirazu.ac.ir}
\affiliation{Department of Physics, School of Science, Shiraz University, Shiraz 71946-84795, Iran}
\author{Mohsen Rahmani Haghighi}
\email{Rahmani.qit@gmail.com}
\affiliation{Department of Physics, School of Science, Shiraz University, Shiraz 71946-84795, Iran}
\begin{abstract}
While local unitary transformations are used for identifying quantum states which are in the same topological class, non-local unitary transformations are also important for studying the transition between different topological classes. In particular, it is an important task to find suitable non-local transformations that systematically sweep different topological classes.  Here, regarding the role of dimension in the topological classes, we introduce partially local unitary transformations namely Greenberger-Horne-Zeilinger (GHZ) disentanglers which reduce the dimension of the initial topological model by a layer-by-layer disentangling mechanism. We apply such disentanglers to two-dimensional (2D) topological quantum codes and show that they are converted to many copies of Kitaev's ladders. It implies that the GHZ disentangler causes a transition from an intrinsic topological phase to a symmetry-protected topological phase. Then, we show that while Kitaev's ladders are building blocks of both color code and toric code, there are different patterns of entangling ladders in 2D color code and toric code. It shows that different topological features of these topological codes are reflected in different patterns of entangling ladders. In this regard, we propose that the layer-by-layer disentangling mechanism can be used as a systematic method for classification of topological orders based on finding different patterns of the long-range entanglement in topological lattice models.
\end{abstract}
\maketitle
\section{Introduction}
Studying equivalence classes under local unitary transformations \cite{1,2,14,4} is an important approach in the classification of quantum phases of matter which is one of the most important problems in condensed matter physics \cite{5,6}. Applying such transformations as local disentanglers to lattice models is an important step of entanglement renormalization which is an important tool for studying critical quantum phases as well as topological quantum phases \cite{a,b,c,d}. In particular, because of non-local order in topological quantum systems \cite{7,8,9,10,11,12,13}, quantum phases in different topological classes can not be transformed to each other by local operations. It implies that different equivalence classes under local unitary transformations correspond to different topological classes \cite{3}.

Among topological quantum systems, the classification of topological quantum codes has attracted much attention due to their applications in quantum computation \cite{15,16,17,18,19}. Toric code model \cite{20} with a $Z_2$ topological order \cite{21} is a quantum memory that is topologically robust against local perturbations \cite{22,23,24,25,26,27}. Another important topological code is color code \cite{17,18,28} with an additional element of color which leads to more computational power compared with the toric code. Local unitary transformations play an important role in characterizing these topological codes. In particular, it has been shown that a 2D color code is locally unitary equivalent to two copies of toric codes \cite{29,30,31,32,33,34,35,zare,zare2}, and therefore, color code has a $Z_2 \times Z_2$ topological order.

Local unitary transformations are also important in understanding the role of dimension in the classification of topological phases. For example, one-dimensional quantum states are topologically trivial because a local unitary transformation converts them to product states like a scissor that breaks a string. Therefore, there is no intrinsic topological order in one-dimensional lattice models \cite{e}. However, it is known that some 1D quantum phases have a non-intrinsic topological order and are named symmetry-protected topological phases. A simple example of such models is the toric code state on a ladder which shows a topological phase protected by a $Z_2 \times Z_2 $ symmetry \cite{f} in the sense that it is not transformed to a product state by local unitary transformations which respect to a $Z_2 \times Z_2 $ symmetry.

On the other hand, topological order is characterized by a long-range entanglement in topological quantum states \cite{z1}. Since local unitary transformations can not remove the long-range entanglement in a topological state, if we consider the space of all quantum states belonging to different topological classes, local unitary transformations correspond to moving along paths towards fixed points in each topological class \cite{3}. However, in order to move between different topological classes, we need non-local unitary transformations to change the pattern of the long-range entanglement. Therefore, it is an important task to find a systematic way for applying non-local transformations to sweep all topological classes.

Here we propose partially local unitary transformations which are local along one particular dimension of the lattice and non-local along other dimensions. We show that it leads to a layer-by-layer disentangling mechanism that induces transitions between different topological classes by reducing the dimension of the initial quantum state. We explicitly introduce such a partially local transformation that we call GHZ disentangler for color code on a hexagonal lattice as well as toric code on a square lattice. By applying such disentanglers to the above 2D topological codes, we convert them to many copies of Kitaev codes on ladders that have symmetry-protected topological phase. Therefore, it implies a transition from intrinsic topological phases to the symmetry-protected topological phase. We also use our results for comparing the entanglement structure of the color code with that of the toric code. In particular, we show that the difference between these important topological quantum codes is reflected in different patterns of the entanglement between Kitaev's ladders. In this regard, we propose that the layer-by-layer disentangling mechanism is an important tool for finding pattern of the long-range entanglement in different topological states which is important for classification of topological orders.

The structure of the paper is as follows: In Sec.(\ref{sec1}), we give an introduction to the toric code, Kitaev ladder, and color code.  In Sec.(\ref{sec2}), we introduce a partially local unitary transformation for color code state on a hexagonal lattice. We show that such a transformation plays the role of a disentangler which converts the color code state to many copies of Kitaev's ladders. In Sec.(\ref{sec3}), we examine our approach for the toric code state and show that it is also converted to many copies of Kitaev's ladders by partially local transformations. Finally, we compare the pattern of long-range entanglement in toric code and color code by considering different patterns of entangling ladders in these topological codes.

\section{Topological quantum codes}\label{sec1}
Toric code (TC) is one of the most pioneer quantum codes \cite{15,19} which can be defined on any arbitrarily oriented lattice with qubits on the edges, see Fig.(\ref{kitaev}). The Hamiltonian corresponding to this code is defined in terms of vertex and plaquette operators $A_v$ and $B_p$
\begin{equation}\label{TM}
H_{TC}=-J\sum_v A_v -J\sum_p B_p
\end{equation}

where $J$ is the coupling energy. $B_p$ and $A_v$ are defined as follows:
\begin{equation}\label{sta}
B_p =\prod_{i\in \partial p} Z_i ~~,~~A_v =\prod_{i\in v}X_i
\end{equation}
where $X$ and $Z$ are Pauli operators, $i\in \ v$ refers to qubits that live on edges incoming to the vertex $v$ and $i\in \partial p$ refers to qubits that live on edges surrounding the plaquette $p$. In Fig.(\ref{kitaev}), we show these operators for three different lattices including a square lattice, a triangular lattice, and a triangular ladder. Since the plaquette and vertex operators are commuted with each other, the ground state of the toric code is simply obtained as follows:
\begin{figure}[H]
\centering
\includegraphics[width=8cm,height=5cm,angle=0]{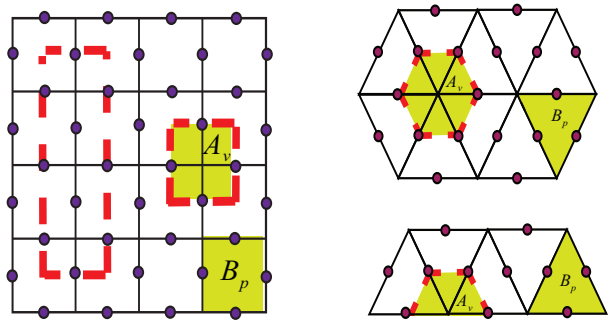}
\caption{Toric code on a square lattice, triangular lattice, and triangular ladder where qubits live in edges. $A_v$ operators are represented by a loop on a dual lattice denoted by red (dashed) lines and $B_p$ operators correspond to plaquettes of the lattice. A product of $A_v$ operators for square lattice is represented by a bigger loop on dual lattice.} \label{kitaev}
\end{figure}
\begin{equation}\label{kitaevstate}
|GS \rangle =\prod_{v}(1+A_v)|0\ra ^{\otimes n}
\end{equation}
where $n$ is the number of qubits and we ignore the normalization factor. On the other hand, since each vertex operator corresponds to a loop on the dual lattice, as shown in Fig.(\ref{kitaev}), each product of the plaquette operators can be represented by configurations of loops. In this regard, the ground state of a toric code is a superposition of all loop configurations of spin down $|1\ra$ on the background of spin ups  $|0\ra$ which is called a loop condensed state. Such a state has a topological order which leads to degeneracy in the ground state when we consider a periodic boundary condition. In particular, there are two topological operators in the form of the product of $X$ operators along non-contractible loops around the torus. Applying such operators in the $|GS\ra$ generates three more ground states of the toric code. In particular, different topological descriptions of the above ground states are the reason for robust degeneracy in the toric code which is important for application as a quantum memory.

The robustness of topological order in the toric code is also understood in terms of local unitary transformations. In particular, topological order is robust against arbitrary local unitary transformations in the sense that the ground state can not be converted to a product state by applying arbitrary local unitaries. On the other hand,  as shown in Fig.(\ref{kitaev}), we can consider Kitaev code on a quasi- one-dimensional lattice such as a ladder which is named Kitaev's ladder. It is shown that  Kitaev's ladder does not have an intrinsic topological order but it is a symmetry-protected topological phase \cite{f}. In particular, while the ground state is converted to a product state under generic local unitary transformations, it is protected under local unitaries that respect  a particular symmetry i. e. $Z_2 \times Z_2$ symmetry.

\begin{figure}[h!]
\centering
\includegraphics[width=7cm,height=7cm,angle=0]{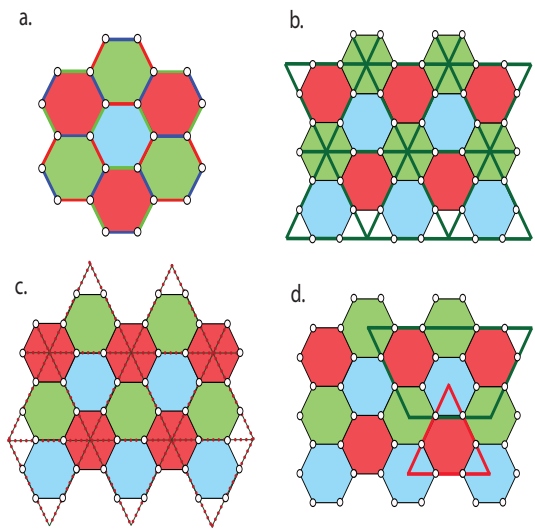}
\caption{a) Color code on a two dimensional hexagonal lattice. Here the qubits live on the vertices of each plaquette and the edges connect the plaquette of the same color. For example, the green edge connects the green plaquettes. b) Red and blue plaquette operators are described by triangles of a green triangular lattice. c) Blue and green plaquette operators are described by triangles of a red triangular lattice. d) A product of plaquette operators in the color code is represented by a loop structure constructed by two different colors.} \label{LoopCC}
\end{figure}

Besides the toric code, color code (CC) is also another topological quantum code in which, qubits live on the vertices of a three-colorable lattice. Adding an extra element of color in this model leads to the emergence of some features which are different from the toric code \cite{17,18}. Here we consider a two-dimensional hexagonal lattice that is colored by three colors, red, blue, and green. As it is shown in Fig.(\ref{LoopCC}-a), the hexagonal lattice is a three-colorable lattice in the sense that no two neighboring plaquettes have the same color. Moreover, the edges are also three colorable where we assign a color to each edge that connects the plaquettes of the same color.

The Hamiltonian corresponding to this code is written as:
\begin{equation}\label{TM}
H_{CC}=-J\sum_p B^x _p -J\sum_p  B^z _p
\end{equation}
where $B_P^X$ and $B_P^Z$ are commuting plaquette operators which are defined as follows:
\begin{equation}\label{sta}
B^z _p =\prod_{i\in p} Z_i ~~,~~ B^x _p =\prod_{i\in p}X_i
\end{equation}
where $i\in p$ refers to all qubits belongings to the plaquette $p$.

Similar to the toric code, the ground state of color code can be written in terms of $X$-type operators as follows, up to a normalization factor:

\begin{equation}\label{ccstate}
|GS \rangle_{cc} =\prod_{p}(1+B^x_p)|0\ra ^{\otimes m}
\end{equation}
where $m$ refers to the number of vertices in the hexagonal lattice. As it is shown in Fig.(\ref{LoopCC}-b,c), we can plot a triangular lattice with edges crossing edges of the hexagonal lattice which have the same color. In this regard, since each triangle of such a lattice corresponds to a hexagonal plaquette of the initial lattice, the corresponding $B^x_p$ operator can be represented by a triangular loop. It implies that there should be a loop representation for the color code state similar to the toric code state. However, it is impossible to represent all $B^x_p$ operators with loops with the same color. For example, while red and blue plaquettes correspond to green triangles Fig.(\ref{LoopCC}-b), for representing green plaquettes we need blue or red triangles Fig.(\ref{LoopCC}-c). In this regard, $\prod_{p}(1+B^x_p)$ in Eq.(\ref{ccstate}) does not lead to a simple loop condensed state. In particular, there are loop structures constructed by different colors similar to what we show in Fig.(\ref{LoopCC}-d).

Existence of loop structures of different colors plays also an important role in degeneracy of the ground state of the color code model. In particular, we have six non-contractible loops with three different colors in two different directions. In this regard and since only two colors of the above non-contractible loops are independent, the color code has a 16-fold degeneracy due to four non-contractible loops of two different colors. Furthermore, it has been shown that the color in the color code leads also to more computational power compared to the toric code where one is able to apply all clifford gates on qubits encoded in the ground state of the color code \cite{17}. In spite of such difference between toric code and color code, it is shown that a local unitary transformation which converts a 2D color code to two copies of the toric code.

\begin{figure}
\centering
\includegraphics[width=7cm,height=5cm,angle=0]{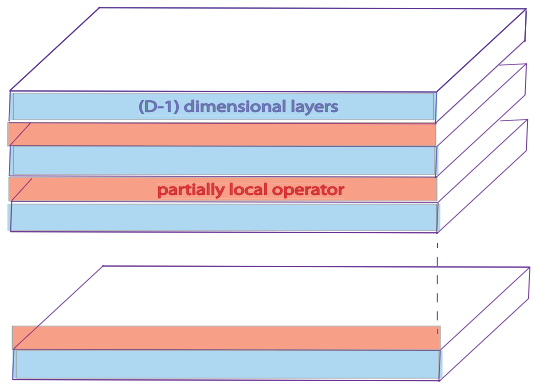}
\caption{(Color online) A schematic of a D-dimensional model where partially local operators are applied between $(D-1)$-dimensional layers.} \label{schematic}
\end{figure}
Here, we would like to emphasize in different topological classes of the above 2D topological codes and quasi-1D Kitaev's ladder. It is a reflection of the role of dimension in the classification of topological phases. In particular, toric code and color code can even be defined on higher-dimensional lattices where different topological properties emerge \cite{bb,42,43,44}. For example, while excitations in the 2D toric code and color code are string-type, in higher dimension excitations correspond to membranes \cite{bb,43}. This important topological property is the reason that higher dimensional version of these codes can be self-corrected \cite{29}. Regarding different topological properties of topological codes in different dimensions, it is clear that there is no local unitary transformation that converts topological codes in different dimensions.

 Here, we propose a systematic way to induce transition between different topological classes corresponding to different dimensions. To this end, for a D-dimensional topological code one can consider a partially local transformation which is applied between two $D-1$ dimensional layers in the sense that while it is non-locally applied to D-1 dimensional layers, it is local in direction orthogonal to the above layers, see Fig.(\ref{schematic}) as a schematic of such a transformation. In the next section, we introduce explicitly such a transformation for 2D color code and show that it converts color code to many copies of the Kitaev's ladders in the sense that it reduces the dimension of the initial topological code by a layer-by-layer disentangling mechanism.

\section{partially local transformations on the color code}\label{sec2}
\begin{figure}
\centering
\includegraphics[width=6cm,height=6cm,angle=0]{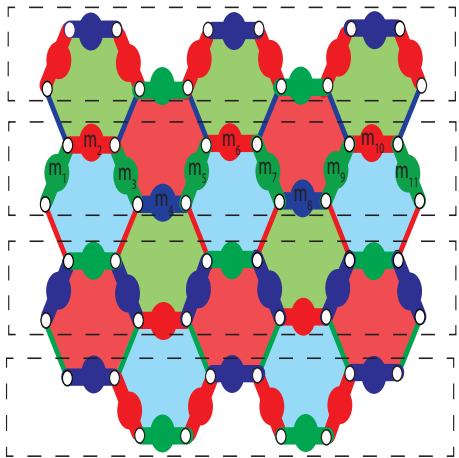}
\caption{(Color online) Partially local transformations are applied to qubits living in non-contractible loops in horizontal directions. Such a transformation corresponds to a change of basis which is represented by GHZ qubits, denoted by circles by three different colors, living on the edges of the lattice along non-contractible loops.} \label{CCladder0}
\end{figure}
In this section, we examine a layer-by-layer disentangling operation for a 2D topological color code and show that it is converted to many copies of Kitaev ladders by a partially local unitary transformation. To this end, consider the color code on the honeycomb lattice. As it is shown in Fig.(\ref{CCladder0}), we consider non-contractible loops along one direction on the lattice where each loop passes from $N$ qubits. There are $N$ numbers of such non-contractible loops on the lattice which cover all qubits of the color code. Then, corresponding to each loop, we introduce an $N$-qubit GHZ basis. To this end, note that the GHZ state on $N$ qubit in the form of $\frac{1}{\sqrt{2}}(|00...0\ra +|11...1\ra)$ is a stabilizer state stabilized by a group of Pauli operators constructed by the following $N$ generators:
\begin{equation}
g^N =\{Z_1 Z_2 , Z_2 Z_3 , ..., Z_i Z_{i+1} , ..., Z_N Z_1 \Omega_x \}
\end{equation}
\begin{figure}
\centering
\includegraphics[width=8cm,height=10cm,angle=0]{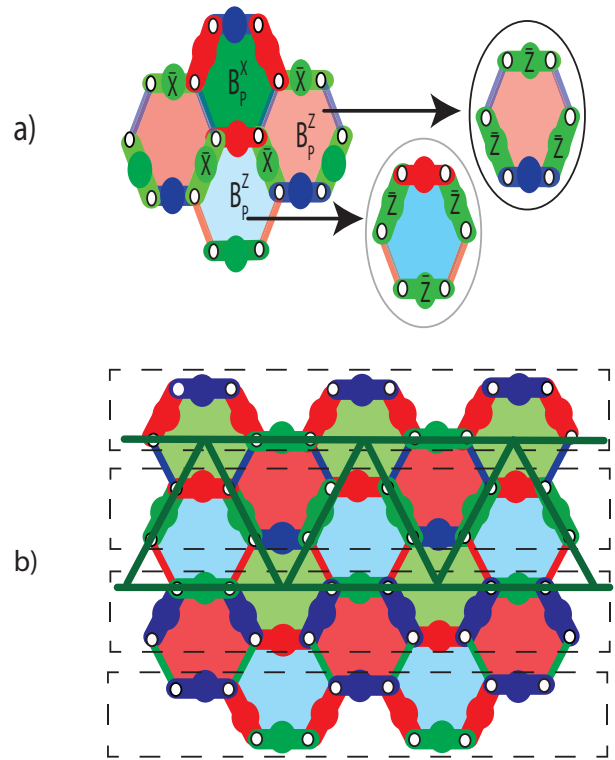}
\caption{(Color online) a) A green $B_p ^x$ operator is converted to a product of $X$ operators on four green GHZ qubits. Blue and red $B_p ^z$ operators are also converted to a product of $Z$ operators on three green GHZ qubits. b) the resultant stabilizers are the same as stabilizers of a Kitaev codes defined on a triangular ladder.} \label{CCladder1}
\end{figure}
where $\Omega_x$ refers to a product of all $X$ operators on qubits belonging to a non-contractible loop and we denote the above generators by $g_1 ,g_2 , ..., g_N $, respectively. Moreover, using the above set of stabilizers, we are also able to construct other $N-1$ GHZ states to have a complete $N$-qubit GHZ basis. For example, the state $\frac{1}{\sqrt{2}}(|00...0\ra -|11...1\ra)$ is stabilized by $g_1$,...,$g_{N-1}$ but the effect of $g_N$ on such state leads to the eigenvalue of $-1$. In the same way, all $N$-qubit GHZ states are defined as eigenstates of $g_1$,...$g_N$ with different eigenvalues. In this regard, we write all $2^N$ GHZ states in the form of:
\begin{equation}
\frac{1}{2^{N/2}} \prod_{i=1}^N (1+(-1)^{m_i}g_i)|++...+\rangle
\end{equation}
where $m_i ={0,1}$ and $m_1$,$m_2$,...,$m_N$ are called GHZ qubits living on edges belonging to each non-contractible loop, see Fig.(\ref{CCladder0}) where we denote GHZ qubits by circles colored by the same color of the corresponding edge. Notice that corresponding to each non-contractible loop there is a GHZ basis and therefore, whole space for $N^2$ qubits on the lattice is spanned by a product of $N$ numbers of the above $N$-qubit GHZ bases.

\begin{figure*}
\centering
\includegraphics[width=12cm,height=6cm,angle=0]{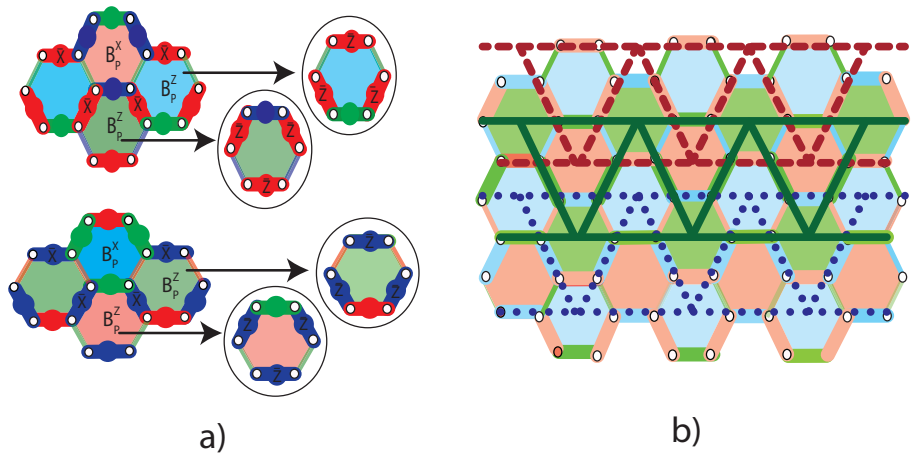}
\caption{(Color online) a)  A red $B_p ^x$ operator beside green and blue $B_p^z$ operators are converted to stabilizers applied on red qubits. A blue $B_p ^x$ operator beside red and green $B_p^z$ operators are converted to stabilizers applied on qubits qubits. b) the resultant stabilizers are the same as stabilizers of  Kitaev codes defined on red and blue triangular ladders. In this regard, color code is constructed by Kitaev's ladders with three different colors which overlap to each other and each disentangler along a non-contractible loop disentangles three successive Kitaev's ladders of the color code.} \label{CCladder2}
\end{figure*}
It is clear that there is a unitary transformation that changes the computational basis to the above GHZ basis. Since such an operator is local in the vertical direction and non-local in the horizontal direction, we call it a partially local unitary transformation. Now, we are going to find the effect of such a transformation on the color code state. Since there is a one-to-one correspondence between a stabilizer state and the group of its stabilizers, we consider the effect of the above transformation on stabilizers of the color code state. Then, we can use the new group of stabilizers to characterize the final quantum state after transformation. To this end, we divide all stabilizers in three sets corresponding to three colors of GHZ qubits. In particular, corresponding to green GHZ qubits, we consider all $B_p^x$ operators corresponding green plaquettes in addition to $B_p ^z$ operators corresponding to red and blue plaquettes which have three green GHZ qubits on their edges, see Fig.(\ref{CCladder1}-a). In the same way, corresponding to the red (blue) GHZ qubits, we consider another stabilizer set including $B_p^x$ operators corresponding the red (blue) plaquette in addition to $B_p ^z$ operators corresponding to blue and green (red and green) plaquettes which have three red (blue) GHZ qubits on their edges, see Fig.(\ref{CCladder2}-a).

We start with transformation on the first set of stabilizers corresponding to the green color.  In particular, consider a $B_p ^x$ stabilizer corresponding to a green plaquette. As shown in Fig.(\ref{CCladder1})-a, there are eight GHZ qubits near a green plaquette including three red GHZ qubits, one blue GHZ qubit, and four green GHZ qubits. To consider the effect of the $B_p ^x$ operator on these eight GHZ qubits, note that each GHZ qubit in the GHZ basis appears in the form of $...(1+(-1)^{m_i} Z_i Z_{i+1})...|++...+\ra$. In this regard, if $B_p ^x$ commutes with $Z_i Z_{i+1}$, we will have $B_p ^x (1+(-1)^{m_i} Z_i Z_{i+1})= (1+(-1)^{m_i} Z_i Z_{i+1}) B_p ^x$  and therefore, the effect of $B_p ^x$ on the GHZ qubit $m_i$ is equivalent to an identity operator. On the other hand, if $B_p ^x$ unticommutes with $Z_i Z_{i+1}$, we will have $B_p ^x (1+(-1)^{m_i} Z_i Z_{i+1})= (1+(-1)^{m_i +1} Z_i Z_{i+1}) B_p ^x$  and therefore, the effect of $B_p ^x$ on the GHZ qubit $m_i$ is equivalent to a logical $X$ operator which shifts $m_i$ to $m_i +1$. In this regard, we consider commutation relation of the green $B_p ^x$ operator with eight operators of $Z_i Z_{i+1}$ corresponding to eight GHZ qubits near the green plaquette $p$. As seen in Fig.(\ref{CCladder1})-a, Since $B_p ^x$ has two qubits common with blue and red edges, it commutes with the corresponding $Z_i Z_{i+1}$. However, it has one qubit common with four green edges and therefore,  it unticommutes with the corresponding $Z_i Z_{i+1}$. In this regard, the effect of $B_p^x$ operator is equivalent to a product of four logical $X$ operators on four green GHZ qubits.

 Now, we consider red and blue plaquettes which have three green GHZ qubits in their edges, and study transformation on the corresponding $B_p ^z$ operators. In particular  note that such a $B_p ^z$ operator has two qubits in common with each green edge and therefore it is equal to a product of $Z_i Z_{i+1}$ operators on three green edges. To consider the effect of this operator on the GHZ basis, we notice that $Z_i Z_{i+1}(1+(-1)^{m_i} Z_i Z_{i+1})=(-1)^{m_i}(1+(-1)^{m_i} Z_i Z_{i+1})$. Therefore, each $Z_i Z_{i+1}$ applied to a green edge is equivalent to a logical $Z$ operator on the corresponding green GHZ qubit. Consequently, the above $B_p ^z$ operators are transformed in to a product of three logical $Z$ operators on three green GHZ qubits around the plaquette $p$ as shown in Fig.(\ref{CCladder1})-a. Interestingly, as shown in Fig.(\ref{CCladder1})-b, the resultant logical $X$-type and $Z$-type stabilizers are the same as the vertex and plaquette operators for a Kitaev code define on a triangular green ladder where green GHZ qubits live on the edges of the ladder. By applying the above transformation to similar stabilizers in other rows of the lattice, we find other green ladders. Importantly, the above ladders are completely separated in the sense that there are no common green GHZ qubits for them.

Transformation for other sets of stabilizers corresponding to blue and red colors is also done in the same way. As shown in Fig. (\ref{CCladder2})-a, consider a $B_p ^x$ stabilizer corresponding to a red (blue) plaquette. Such an operator anticommutes with four red (blue) GHZ qubits and therefore, it is equal to a product of four logical $X$ operators on red (blue) GHZ qubits. $B_p ^z$ operators corresponding to green and blue (green and red) plaquettes are also equal to the product of three logical $Z$ operators on three red (blue) GHZ qubits as shown in Fig.(\ref{CCladder2})-a. Such stabilizers are also represented by red (blue) ladders and the resultant stabilizers are the stabilizers of Kitaev codes defined on red (blue) ladders, see Fig.(\ref{CCladder2}-b).

In this regard, while in the color code state, all qubits are entangled, logical qubits in the GHZ basis are disentangled where the resultant state is a tensor product of Kitaev states on ladders with three different colors. In other words, Kitaev's ladders are building blocks of the color code and the partially local unitary transformation plays the role of a layer-by-layer disentangler which separates different layers of the color code states. Regarding symmetry protected topological phase of the Kitaev's ladder, our result implies a transition from an intrinsic topological phase to a symmetry-protected topological phase. On the other hand, notice that the non-local nature of disentanglers has led to change of pattern of the long-range entanglement in the initial state. Therefore, the pattern of entangling ladders in the color code is in fact a simple picture of pattern of the long-range entanglement in this topological quantum code.

\section{Disentangling toric code to Kitaev's ladders}\label{sec3}
\begin{figure}
\centering
\includegraphics[width=6cm,height=6cm,angle=0]{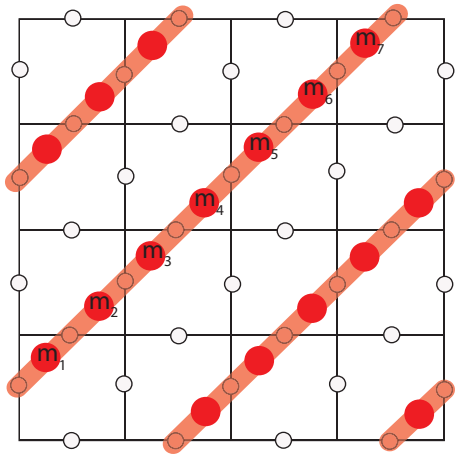}
\caption{(Color online) Partially local unitary operators are applied to diagonal line denoted by orange color. Such a transformation correspond to a change of basis which is represented by GHZ qubits, denoted by circles living along the diagonal lines. } \label{TCladder0}
\end{figure}
\begin{figure}
\centering
\includegraphics[width=8cm,height=8cm,angle=0]{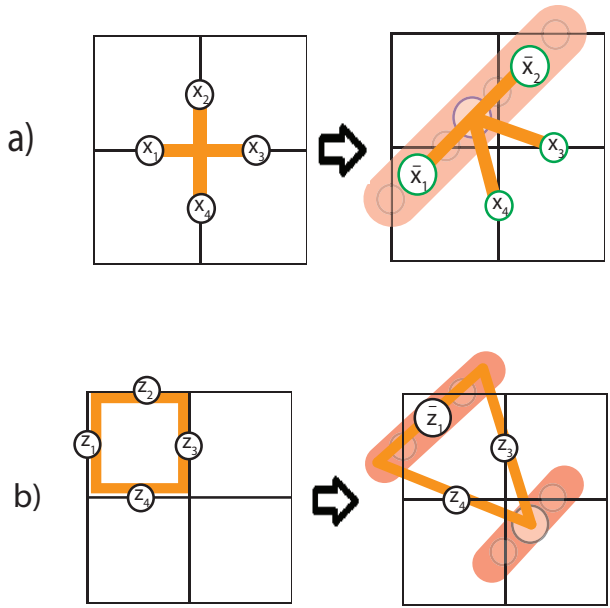}
\caption{(Color online) a) An $A_v$ operator is converted to a product of $X$ operators on two GHZ qubits and two initial qubits. b) A $B_p$ operator is converted to a product of $Z$ operators on one GHZ qubit and two initial qubits. } \label{TCladder1}
\end{figure}

Our layer-by-layer disentangling method can be applied to other topological models. It is in particular important for comparing the patterns of the long-range entanglement for topological states in different topological classes. In this section, we show that there is another pattern of partially local transformations which converts the toric code model to many copies of Kitaev's ladders. To this end, as shown in Fig.(\ref{TCladder0}), we study Toric code on a square lattice and consider diagonal lines on the square lattice which cross qubits along non-contractible loops. Corresponding to half of these lines, we define GHZ bases in the sense that two neighboring qubits $i$ and $i+1$ along the line are mapped to a GHZ qubit $m_i$ living between them i. e. $(1+(-1)^{m_i}Z_i Z_{i+1})$. It would be a partially local transformation because it is local in the direction which is orthogonal to the diagonal lines. Next, we consider the effect of such a partially local transformation on stabilizers of the toric code. As shown in Fig.(\ref{TCladder1}-a), a vertex operator $A_v = X_1 X_2 X_3 X_4$ unticommutes with $Z_i Z_{i+1}$s corresponding to two GHZ qubits corresponding to qubits $1$ and $2$. Therefore $X_1$ and $X_2$ are converted to two logical operators $\bar{X}_1$ and $\bar{X}_2$ while two original operators $X_3$ and $X_4$ remain unchanged. Consequently, the original operator is converted to a four-local operator including two logical $X$ operators and two initial $X$ operators which are the same as the $X$-type stabilizer of a Kitaev code on the triangular ladder. On the other hand, for a plaquette operator $B_p ^z =Z_1 Z_2 Z_3 Z_4 $ shown in Fig.(\ref{TCladder1}-b), while $Z_3$ and $ Z_4$ remain unchanged, the effect of $Z_1 Z_2$ on the GHZ basis is equal to a logical operator  $\bar{Z}_1$ because $Z_1 Z_2 (1+(-1)^{m_1} Z_1 Z_2)=(-1)^{m_1}(1+(-1)^{m_1}Z_1 Z_2)$. Therefore, the initial plaquette operator is converted to a three-local $Z$-type stabilizer including one logical operator $\bar{Z}_1$ and two original operators $Z_3$ and $Z_4$ which is the same as the $Z$-type stabilizer of the Kitaev code on triangular ladder.

\begin{figure*}
\centering
\includegraphics[width=17cm,height=6cm,angle=0]{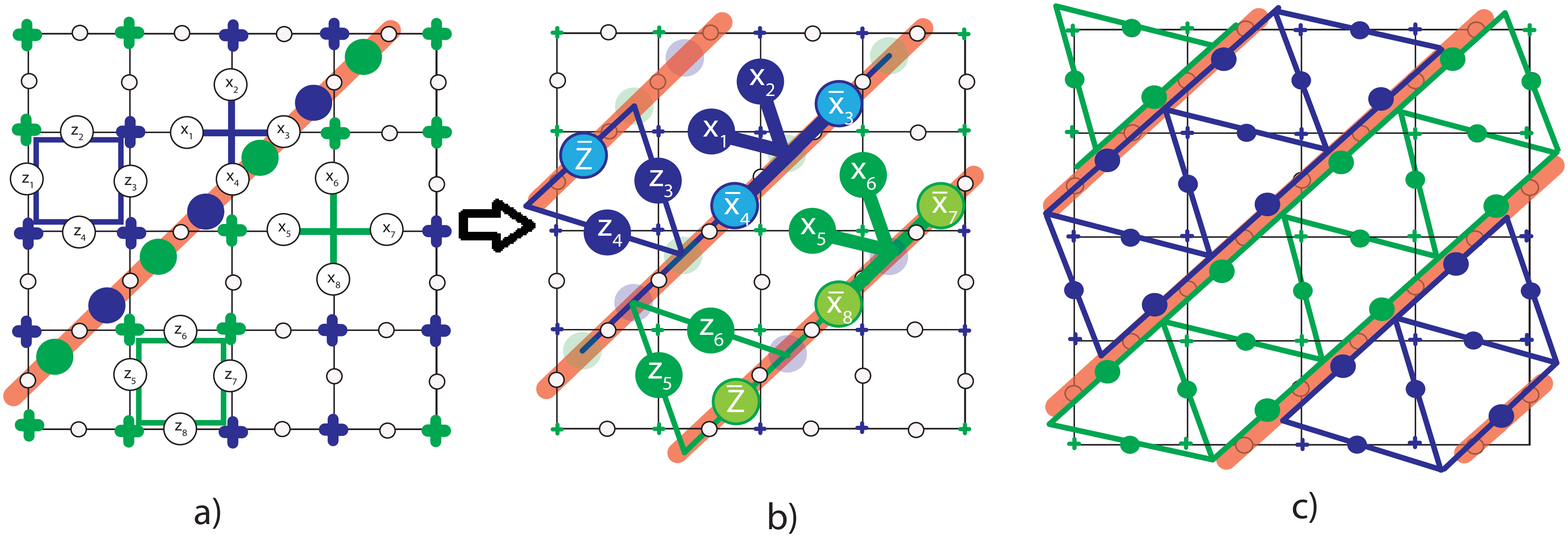}
\caption{(Color online) a) We divide vertices of the square lattice into two sets denoted by blue and green colors. A GHZ qubit living into a square plaquette is also colored by blue or green color if most vertices of that plaquette are blue or green, respectively. b) $A_v$ operators corresponding to blue or green vertices are converted to logical $X$-type stabilizers on blue or green GHZ qubits, respectively. $B_p$ operators are also converted to logical $Z$-type stabilizers on blue or green GHZ qubits. c) The logical stabilizers correspond to blue and green ladders which are decoupled and therefore, the initial toric code is converted to many copies of Kitaev's ladders.} \label{TCladder3}
\end{figure*}

Next, we apply the above transformation to all plaquette and vertex operators of the toric code. As it is shown in Fig.(\ref{TCladder3}-a), we divide all vertices of the square lattice into two different sets denoted by blue and green colors. Then we color also all GHZ qubits with blue and green colors in the sense that a GHZ qubit living in a plaquette is colored by blue (green) if most of the vertices of that plaquette are blue (green). As shown in Fig.(\ref{TCladder3}-b), by such a division of vertex operators, the blue and green vertex operators are converted to $X$-type operators living in blue and green GHZ qubits, respectively. In other words, blue and green GHZ qubits are decoupled due to transformation. In the same way, plaquette operators are also converted to two sets of $Z$-type operators on blue and green GHZ qubits which are decoupled. Finally,  applying the above transformation on all qubits generates blue and green Kitaev's ladders which are decoupled, see Fig.(\ref{TCladder3}-C). Consequently, similar to color code, Kitaev's ladders are building blocks of the toric code and the partially local transformation plays the role of a layer-by-layer disentangler which separates diagonal ladders from the toric code.

On the other hand, it is important to compare patterns of the entanglement between ladders for the toric code with that for the color code. Regarding Fig.(\ref{TCladder3}-C), toric code is constructed by entangling Kitaev's ladders which are inserted near each other in a side-by-side pattern. However, as shown in Fig.(\ref{CCladder2}-b), color code has a different structure in the sense that it is constructed by Kitaev's ladders which overlap with each other and entanglers are applied to three successive Kitaev's ladders. In particular, we represent the above three Kitaev ladders with three different colors corresponding to three colors in the color code. In this regard, the above structure is a reflection of the role of color in the difference between features of the color code and the toric code. This result shows that by using the layer-by-layer disentangling mechanism, we are in fact able to find the pattern of long-range entanglement in the above topological quantum states. In other words, different patterns of the entanglement between layers for color code and toric code correspond to different patterns of the long-range entanglement.

\section{Conclusion}
We proposed a layer-by-layer disentangling mechanism as a systematic way for reducing dimension in topological lattice models. Since such non-local operations can change the pattern of long-range entanglement in topological states, it was expected that the above mechanism leads to a transition between different topological classes. We showed that there is such a transition from 2D topological codes with intrinsic topological order to Kitaev's ladders with symmetry-protected topological order. Furthermore, we showed that there are different patterns of the entanglement between ladders for color code and toric code. Therefore, we concluded that different patterns of the entanglement between layers are related to different topological features of the above quantum codes. In other words, different patterns of entanglement between layers correspond to different patterns of the long-range entanglement which are finger prints of different topological classes. As a concluding remark, we propose that the above idea can be used for classification of topological orders in different lattice models with different dimensions. For example, we expect that different topological classes in D dimensional models are distinguished by different patterns of entanglement between layers when we reduce the dimension of the model step by step to convert the initial model to many copies of one dimensional models. In this regard, we would be able to classify different topological orders corresponding to different patterns of entanglement between layers.


\begin{thebibliography}{99}
\bibitem{1}
B. Kraus, Local unitary equivalence of multipartite pure states, Phys. Rev. Lett. 104,020504 (2010).
\bibitem{2}
B. Kraus, Local unitary equivalence and entanglement of multipartite pure states, Phys. Rev. A 82, 032121 (2010).
\bibitem{14}
B. Liu, J. L. Li, X. Li, and C. F. Qiao, Local unitary classification of arbitrary dimensional multipartite pure states, Phys. Rev. Lett. 108, 050501 (2012).
\bibitem{4}
M. Grassl, M. Rotteler, and T. Beth, Computing local invariants of quantum-bit systems, Phys. Rev. A 58, 1833 (1998).
\bibitem{5}
S. Sachdev, Quantum Phase Transition, Cambride University Press.(2nd ed.). ISBN 978-0-521-51468-2.
\bibitem{6}
Carr, Lincoln D. Understanding Quantum Phase Transitions. CRC Press. ISBN 978-1-4398-0251-9.
\bibitem{a}
G. Vidal, Entanglement renormalization, Phys. Rev. Lett. 99,220405 (2007).
\bibitem{b}
G. Vidal, J. I. Latorre, E. Rico, A. Kitaev, Entanglement in quantum critical phenomena, Phys. Rev. Lett. 90:227902,(2003).
\bibitem{c}
M. Aguado, G. Vidal, Entanglement renormalization and topological order, Phys. Rev. Lett. 100, 070404 (2008).
\bibitem{d}
R. N. C. Pfeifer, G. Evenbly, G. Vidal, Entanglement renormalization, scale invariance, and quantum criticality, Phys. Rev. A 79(4), 040301(R) (2009).
\bibitem{7}
L. D. Landau, on the Theory of Phase Transitions, Phys. Z. Sowjetunion, Vol. 11, No. 26, P. 545 (1937)
\bibitem{8}
 L. D. Landau, the Theory of Phase Transitions, Nature 138, 840-841 (1936)
\bibitem{9}
 X.-G. Wen, Topological order and excitation in fractional quantum Hall states, Adv. Phys. 44, 405 (1995)
\bibitem{10}
X.-G. Wen, Vacuum degeneracy of chiral spin state in compactified spaces, Phys. Rev. B, 40, 7387 (1989).
\bibitem{11}
X.-G. Wen, Topological order in rigid states, Int. J. Mod. Phys. B. 4(2):239 (1990).
\bibitem{12}
X.-G. Wen and Q. Niu, Ground-state degeneracy of the fractional quantum Hall states in the presence of a random potential and on high-genus Reimann surfaces, Phys. Rev. B 41, 9377 (1990).
\bibitem{13}
X.-G. Wen, Quantum field theory of many body systems- from the origin of sound to an origin of light and electrons, Oxford univ. Press, Oxford (2004).
\bibitem{3}
X. Chen, Z,-C. Gu, and X.-G. Wen, Local unitary transformation,long-range quantum entanglement, wave function renormalization and topological order, Phys. Rev. B 82, 155138 (2010).
\bibitem{15}
E. Dennis, A. Kitaev, A. Landahl and J. Preskill, Topological quantum memory, J. Math. Phys., vol. 43, p. 4452 (2002).
\bibitem{16}
A. Y. Kitaev, Faulat-tolerant quantum computation by anyons, Ann. Phys., vol. 303, p.160502 (2007).
\bibitem{17}
H. Bombin and M. A. Martin-Delgado, Topological quantum distillation, Phys. Rev. Lett., vol. 98, p. 160502 (2007).
\bibitem{18}
H. Bombin and M. A. Martin-Delgado, Topological computation without braiding, Phys. Rev. Lett. 98, 160502 (2007).
\bibitem{19}
B. J. Brown, D. Loss, J. K. Pachos, C. N. Self, J. R. Wootton, Quantum memories at finite temperature, Rev. Mod. Phys. 88, 045005 (2016).
\bibitem{20}
A. Yu. Kitaev, Fault- tolerant quantum computation by anyons, Ann. Phys. 303, 2 (2003).
\bibitem{21}
C. L. Kane and E. J. Male, $Z_2$ Topological order and the quantum spin Hall effect, Phys. Rev. Lett., vol. 95, no. 14 (2005).
\bibitem{22}
A. Jamadagni, H. Weimer, A. Bhattacharyya, Robustness of topological order in the toric code with open boundaries, Phys. Rev. B 98, 235147 (2018)
\bibitem{23}
S. Trebst, P. Werner, M. Troyer, K. Shtengel, and C. Nayak, Breakdown of a topological phase: quantum phase transition in a loop gas model with tension, Phys. Rev. Lett. 98, 070602 (2007).
\bibitem{24}
A. Hamma and D. A. Lidar, Adiabatic preparation of topological order, Phys. Rev. Lett. 100, 030502 (2008).
\bibitem{25}
S. Dusuel, M. Kamfor, R. Orus, K. P. Schmidt, and J. Vidal, Robustness of a perturbed topological phase, Phys. Rev. Lett. 106, 107203 (2011)
\bibitem{26}
C. Castelnovo and C. Chamon, Quantum topological phase transition at the microscopic level, Phys. Rev. B 77, 054433 (2008).
\bibitem{27}
M. H. Zarei, Robustness of topological quantum codes: Ising perturbation, Phys. Rev. A 91, 022319 (2015).
\bibitem{28}
H. Bombin and M. A. Martin-Delgado, Quantum measurements and gates by code deformation, J. Phys. A: Math. Theor: 42.9 (2009).
\bibitem{29}
A. Kubica, B. Yoshida, and F. Pastawski, Unfolding the color code, New J. Phys. 17 083026 (2015).
\bibitem{30}
A. B. Aloshious and P. K. Sarvepalli, Projecting three-dimensional color codes onto three-dimensional toric codes, Phys. Rev. A 98, 012302 (2018).
\bibitem{31}
A. Kubica and N. Delfosse, Efficient color code decoders in $d\leqslant 2$ dimensions from toric code decoders, arXiv preprint arXiv:1905.07393 (2019).
\bibitem{32}
H. Bombin, G. Duclos-Cianci, and D. Poulin, Universal topological phase of 2D stabilizer codes, New J. Phys., 14:073048 (2012).
\bibitem{33}
N. Delfosse, Decoding color codes by projecting onto surface codes. Phys. Rev. A, 89:012317 (2014).
\bibitem{34}
A. B. Aloshious and P. K. Sarvepalli, Erasure decoding of two-dimensional color codes, Phys. Rev. A 100, 042312 (2019).
\bibitem{35}
A. Bhagoji and P.sarvepalli, Equivalence of 2d color codes (without translational symmetry) to surface codes, 2015 IEEE International Symposium on Information Theory (ISIT), (1):1109-1113, (2015).
\bibitem{zare}
M. H. Zarei, Quantum phase transition from $Z_2 \times Z_2$ to $Z_2$ topological order, Phys. Rev. A, 93(4), 042306 (2016).
\bibitem{zare2}
M. H. Zarei, J. Abouie, Topological line in frustrated Toric code models, Phys. Rev. B, 104(11), 115141 (2021).
\bibitem{e}
X. Chen, Z.-C Gu and X. -G. Wen, Classification of Gapped Symmetric Phases in 1D Spin Systems, Phys. Rev. B 83, 035107 (2011).
\bibitem{f}
A. Langari, A. Mohammad-Aghaei and R. Haghshenas, Quantum phase transition as an interplay of Kitaev and Ising interactions, Phys. Rev. B 91, 024415 (2015).
\bibitem{z1}
X.-G. Wen, Topological order: from long-range entangled quantum matter to an unification of light and electrons, ISRN Condensed Matter Physics, vol. 2013, Article ID 198710, 20 pages, (2013).
\bibitem{bb}
H. Bombin and M. A. Martin-Delgado, Exact Topological Quantum Order in $D = 3$ and Beyond:
Branyons and Brane-Net Condensates, Phys. Rev. B 75:075103, (2007).
\bibitem{42}
H. Bombin, Gauge color codes: optimal transversal gates and gauge fixing in topological stabilizer codes, New J. Phys. 17 083002 (2015).
\bibitem{43}
A. Hamma, P. Zanardi, and X.-G Wen, String and membrane condensation on three-dimensional lattices, Phys. Rev. B 72, 035307 (2005).
\bibitem{44}
C. Castelnovoand C. Chamon, Topological order in a three-dimensional toric code at finite temprature, Phys. Rev. B 78, 155120 (2008).
\bibitem{45}
H. Bombin, R. W. Chhajlany, M. Horodecki, M. A. Martin-Delgado, Self-correcting quantum computers, New J. Phys. 15(5), 055023 (2013).
\end{thebibliography}
\end{document}